\documentclass[twocolumn,prb,showpacs]{revtex4}

\usepackage{graphicx}
\usepackage{bm}

\begin{document}

\title{Differences Between Hole and Electron
Doping \\ of a Two-Leg CuO Ladder}

\author{S.~Nishimoto}
\email[E-mail: ]{Satoshi.Nishimoto@Physik.Uni-Marburg.De}
\author{E.~Jeckelmann}
\email[E-mail: ]{Eric.Jeckelmann@Physik.Uni-Marburg.De}
\affiliation{Philips-Universit\"{a}t  Marburg,
Fachbereich Physik,
D-35032 Marburg, Germany} 

\author{D.J.~Scalapino}
\email{djs@vulcan2.physics.ucsb.edu}
\affiliation{Department of Physics, University of California\\ 
Santa Barbara, California 93106-9530}

\date{\today}

\begin{abstract}

Here we report results of a
density-matrix-renormalization-group (DMRG)
calculation of the charge, spin, and pairing
properties of a two-leg CuO Hubbard
ladder.  The outer oxygen atoms as well as
the rung and leg oxygen atoms are included
along with near-neighbor and
oxygen-hopping matrix
elements.  This system allows us to study the
effects of hole and electron doping on a
system which is a charge transfer insulator
at a filling of one hole per Cu and
exhibits power law, $d$-wave-like pairing
correlations when doped. In particular, we
focus on the differences between doping with
holes or electrons.

\end{abstract}

\pacs{71.10.Fd,74.20.Mn,71.10.Li}
\maketitle


\section{Introduction}

Two-leg CuO ladder materials are known to
exhibit some of the basic
physical properties of the high $T_c$ cuprates.
Undoped two-leg ladder materials
are found to be spin-gapped, charge transfer
insulators~\cite{HATB91,AHTIK94} and
superconductivity has been observed in
hole-doped ladders under high 
pressure.~\cite{Ueh96,Iso98,May98}
Moreover, ladder models have
proved amenable to numerical studies, allowing
one to explore the relationship between the
parameters in the model and physical
properties such as the charge and spin gaps of
the insulating state and the pairing
correlations in the doped
state.~\cite{Dag99,SW01}
Here we continue this type
of theoretical study by examining the
properties of a CuO ladder using a
density-matrix-renormalization-group (DMRG)
analysis.~\cite{Whi93} We will focus on the
differences between hole and electron doping.

\begin{figure}
\includegraphics[width=6cm]{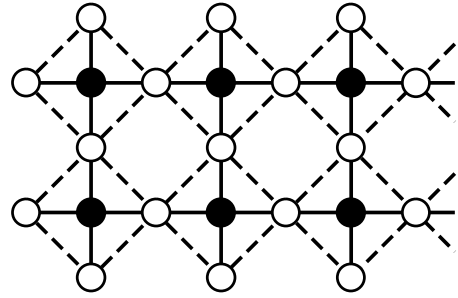}
\caption{\label{fig1} Schematic lattice structure of a two-leg 
CuO ladder.  Here the solid circles represent 
Cu$(d_{x^2-y^2})$ 
orbitals and the open circles represent O$(p_y)$ 
orbitals on the 
rungs and O$(p_x)$ orbitals on both legs. 
The orbital phases are chosen so that there is 
a hopping matrix element $t_{pd}$ between 
nearest-neighbor Cu and O sites represented by the solid lines
and a hopping matrix element $t_{pp}$  between
nearest-neighbor O sites represented by the 
dashed lines. }
\end{figure}

The geometry of the two-leg ladder which we will
study is illustrated in Fig.~\ref{fig1}. The Cu
sites are characterized by a $d_{x^2-y^2}$
orbital, the rung O sites by a $p_y$
orbital, and the leg O sites by a $p_x$
orbital. The hole Hamiltonian for this model is
given by
\begin{eqnarray}
H &=& 
\Delta_{pd} \sum_{i\sigma}
p^\dagger_{i\sigma} p_{i\sigma}
- t_{pd} \sum_{\langle
ij\rangle\sigma} \left(d^\dagger_{i\sigma}
p_{j\sigma} + p^\dagger_{j\sigma}
d_{i\sigma}\right) \nonumber \\
&& - t_{pp} \sum_{\langle ij\rangle\sigma}
\left(p^\dagger_{i\sigma}p_{j\sigma} +
p^\dagger_{j\sigma}
p_{i\sigma}\right) \nonumber \\
&& +U_d \sum_i d^\dagger_{i\uparrow}
d_{i\uparrow} d^\dagger_{i\downarrow}
d_{i\downarrow}
+ U_p \sum_i p^\dagger_{i\uparrow}
p_{i\uparrow} p^\dagger_{i\downarrow}
p_{i\downarrow} 
\label{hamiltonian}
\end{eqnarray}
Here $t_{pd}$ is the one-hole     
hopping matrix element between nearest-neighbor
Cu and O sites and the first sum in 
Eq.~(\ref{hamiltonian})
is over all nearest-neighbor sites.
We will assume for simplicity that
$t_{pd}$ has the same value between all the
Cu and O sites. The second term in
Eq.~(\ref{hamiltonian}) sums over all
nearest-neighbor pairs of O sites and
$t_{pp}$ is the hopping matrix elements
between these sites.
We have chosen the phases of the orbitals
such that the signs of the hopping matrix
elements are constant and with the minus
sign convention of Eq.~(\ref{hamiltonian}),
$t_{pd}>0$ and $t_{pp}\geq 0$.
The energy difference between the
O and Cu sites is
$\Delta_{pd}=\epsilon_p-\epsilon_d > 0$ and $U_d$,
and $U_p$ are the on-site Coulomb energies for
Cu and O, respectively.  
In this model an undoped ladder corresponds to 
a density of one hole per Cu site.
We will work in units where $t_{pd}=1$
and take as typical values $U_d=8$ and $U_p=3$,
throughout. 

Two of the authors previously studied a similar 
model~\cite{JSW98} in which the outer O sites were
missing and only the near-neighbor Cu-O
hopping $t_{pd}$ was present. While this
simplified form allowed one to compare a
charge-gap-insulator model with the traditional
one-band Hubbard and $t$-$J$ models, the
absence of the outer O and the $t_{pp}$ 
hoppings altered the electronic
structure.~\cite{ALJP95,Mul98} 
In addition,
the electron-doped pairing response of the
model appeared to be quite different from the
hole-doped behavior.  Here, for the more
realistic structure shown in Fig.~\ref{fig1}, we return
to the study of the two-leg CuO ladder 
in which the charge-transfer nature
of the insulating state is treated along with
the effects of the oxygen $t_{pp}$ hopping matrix
elements. We also will study longer ladders 
with up
to $32\times2$ Cu atoms corresponding to 226
total O and Cu sites.  

We calculate the static properties of the 
model~(\ref{hamiltonian}) numerically 
using the DMRG method.~\cite{Whi93}  
With this approach we obtain accurate 
ground state energies and expectation values
(such as correlation functions)
for a fixed number of holes of each spin.
In the DMRG calculations, open-end
boundary conditions are used 
so that when we discuss
local quantities they will actually be averaged
over the ladder and
correlation functions will
be calculated using distances taken about the
midpoint.
We have used up to $m=2000$ density matrix 
eigenstates to build the DMRG basis.
Using an extrapolation of DMRG ground state energies
to vanishing discarded weight,~\cite{Bon00} 
we have obtained ground state energies and gaps which are
accurate to parts in $10^{-3}t_{pd}$. 
Although the largest source of errors in our calculations are
finite size effects,
we have been able to extrapolate a number of gap
measurements to an infinite-size ladder.

In Sec.~\ref{sec:undoped} we discuss our
results for a filling of one hole
per Cu, which corresponds to the undoped
ladder. 
Our main interest is in the effect of 
the site energy difference $\Delta_{pd} =
\epsilon_p-\epsilon_d$ and the 
oxygen hopping parameter $t_{pp}$. 
We will examine the charge and
spin magnitudes on the various sites
and the dependence of the charge
and spin gaps on $\Delta_{pd}$ and 
$t_{pp}$.  When
$U_d$ is large compared with $\Delta_{pd}$
and $\Delta_{pd} \agt 2 t_{pp}$, we find as
expected that the ladder is a charge transfer
insulator with a spin gap. By comparing the
low-lying spin states of the CuO model with a
two-leg Heisenberg model, we extract effective
rung and leg exchange interactions. 
In Sec.~\ref{sec:doped},
we examine the charge and spin
magnitude distributions for the doped case and  
find that, for typical parameters characteristic
of the cuprates, the doped holes tend to be 
spread
out on the O lattice while doped electrons tend
to be localized on the Cu sites.  We also study
the effect of doping on the effective exchange
interactions and on the local spin-spin
correlations. 
We then turn to a study of the pairing
correlations for both the hole-doped and the
electron-doped ladder. We find a power law
behavior with negative ``$d$-wave-like'', Cu rung-leg
pair correlations for both hole and electron
doping. However, we find that the internal structure
of a pair depends upon whether the system is 
electron- or hole-doped.
In Sec.~\ref{sec:conclusion},
we conclude with some further comments
relating these results to the general problem
posed by the cuprate materials.

\section{The Undoped Ladder \label{sec:undoped}}

\begin{figure}
\includegraphics[width=6cm]{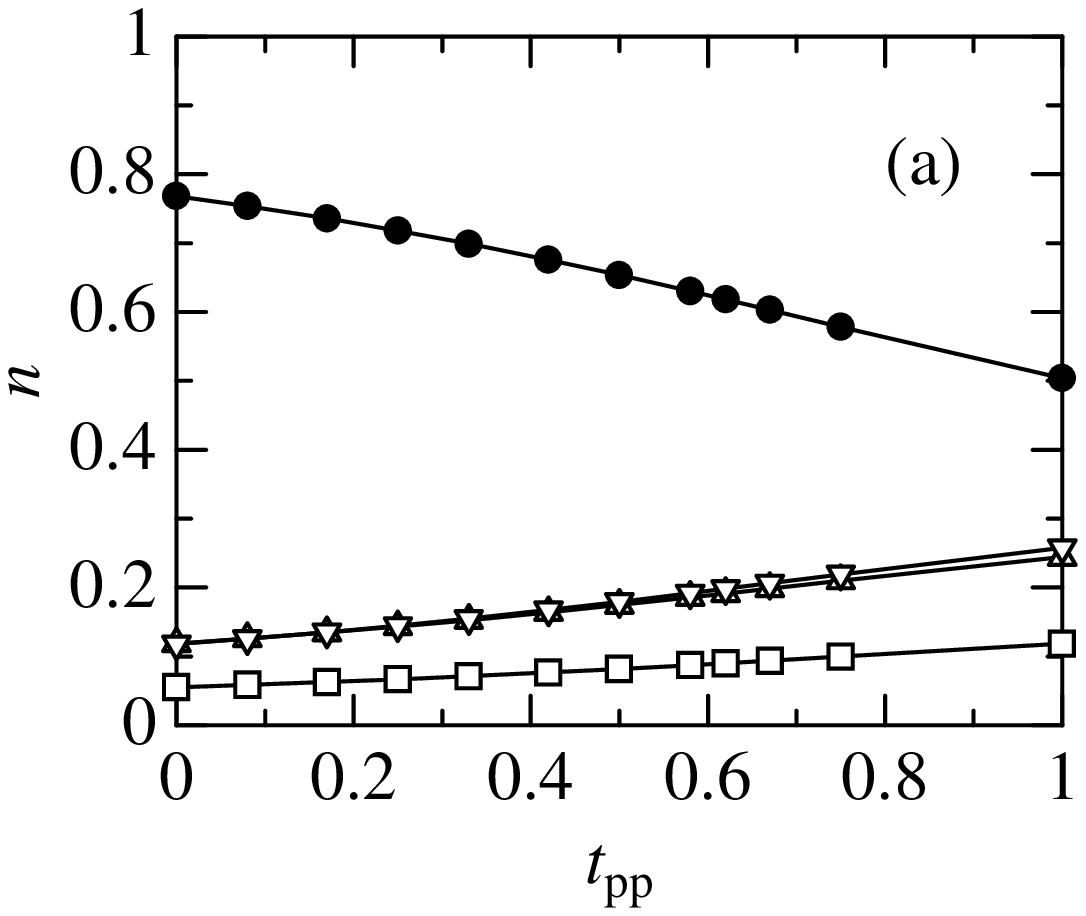}
\includegraphics[width=6cm]{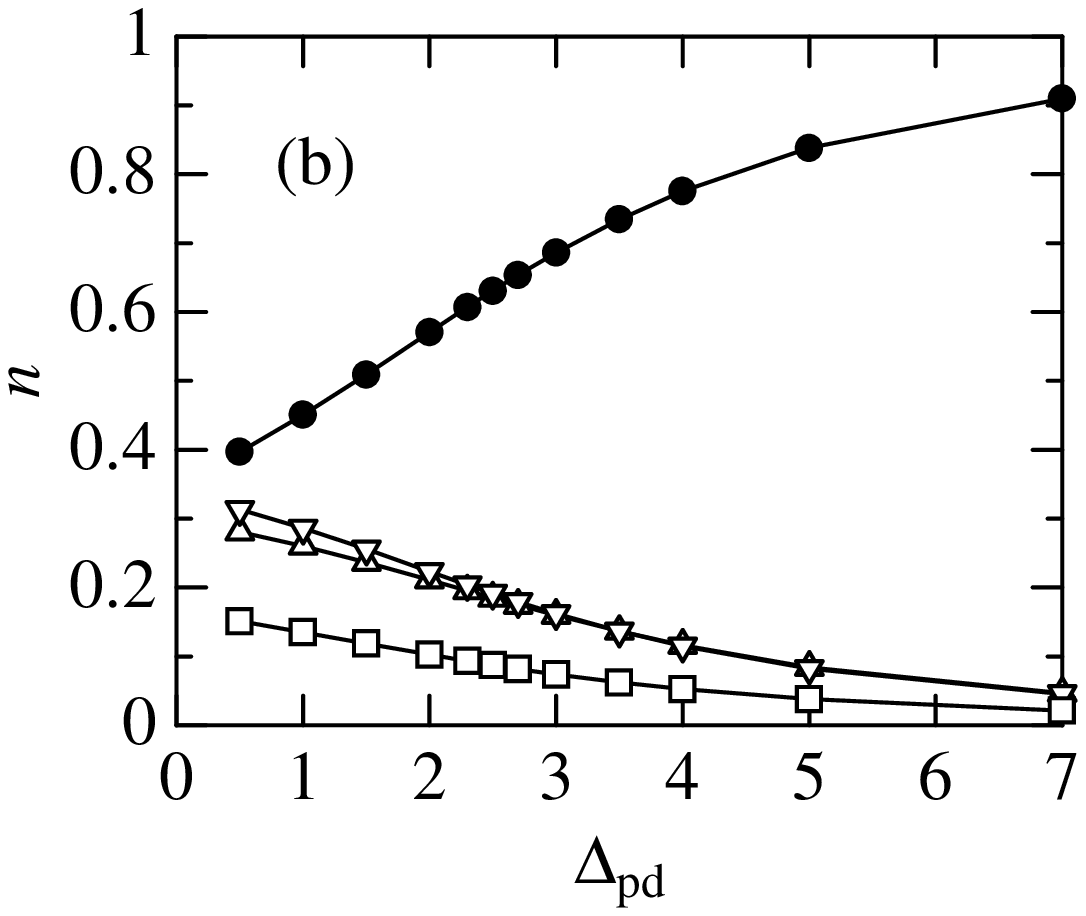}
\caption{\label{fig2} Hole density $\langle n_{\text{Cu}} \rangle$
and $\langle n_{\text{O}} \rangle$
on the various sites
of an undoped ladder with $8\times 2$ Cu atoms:
Cu sites (solid circles), rung O($p_y$) sites (up-triangles), leg 
O($p_x$) sites (down-triangles),
and outer O sites (squares) versus (a) $t_{pp}$ for $\Delta_{pd}=2.7$
and (b) $\Delta_{pd}$ for $t_{pp}=0.5$.}
\end{figure}

We first investigate the properties of an undoped ladder.
For local properties such as the charge and
spin magnitude distributions on
the sites and the spin correlations between sites, a ladder 
containing $8\times 2$ Cu sites is sufficiently long. 
In Fig.~\ref{fig2} we show the average ground state hole 
occupation of Cu sites 
$\langle n_{\text{Cu}} \rangle = 
\langle d^\dagger_{i\uparrow} d_{i\uparrow} + d^\dagger_{i\downarrow}
d_{i\downarrow} \rangle$ and of the various O sites
$\langle n_{\text{O}} \rangle =
\langle p^\dagger_{i\uparrow} p_{i\uparrow} + p^\dagger_{i\downarrow}
p_{i\downarrow} \rangle$  as a function of $t_{pp}$ and $\Delta_{pd}$.
In Fig.~\ref{fig2}(a) the hole occupation is shown for 
$\Delta_{pd}=2.7$ versus $t_{pp}$. 
As $t_{pp}$ increases the hole occupation on the Cu sites decreases 
and that on the O sites increases. 
In Fig.~\ref{fig2}(b), the hole occupation is
shown for $t_{pp}=0.5$ versus $\Delta_{pd}$.  Here as
$\Delta_{pd}$ increases, the hole occupation on the
Cu sites increases as one would expect.
The hole occupation on the outer O sites is typically 
about half that on
the other O sites. For $t_{pp}=0.5$ and
$\Delta_{pd}=3$, the holes are approximately
70\% on the Cu sites and 30\% on the O 
sites. 

\begin{figure}
\includegraphics[width=6cm]{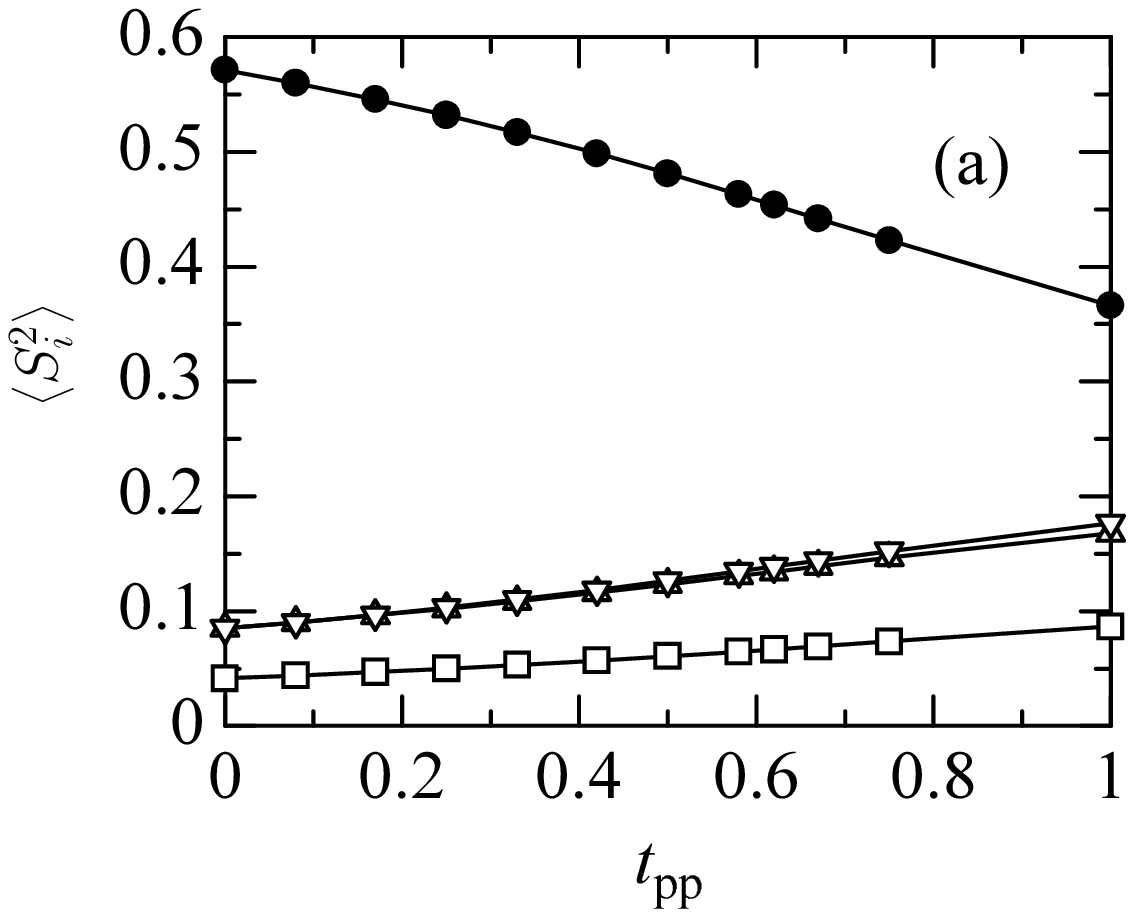}
\includegraphics[width=6cm]{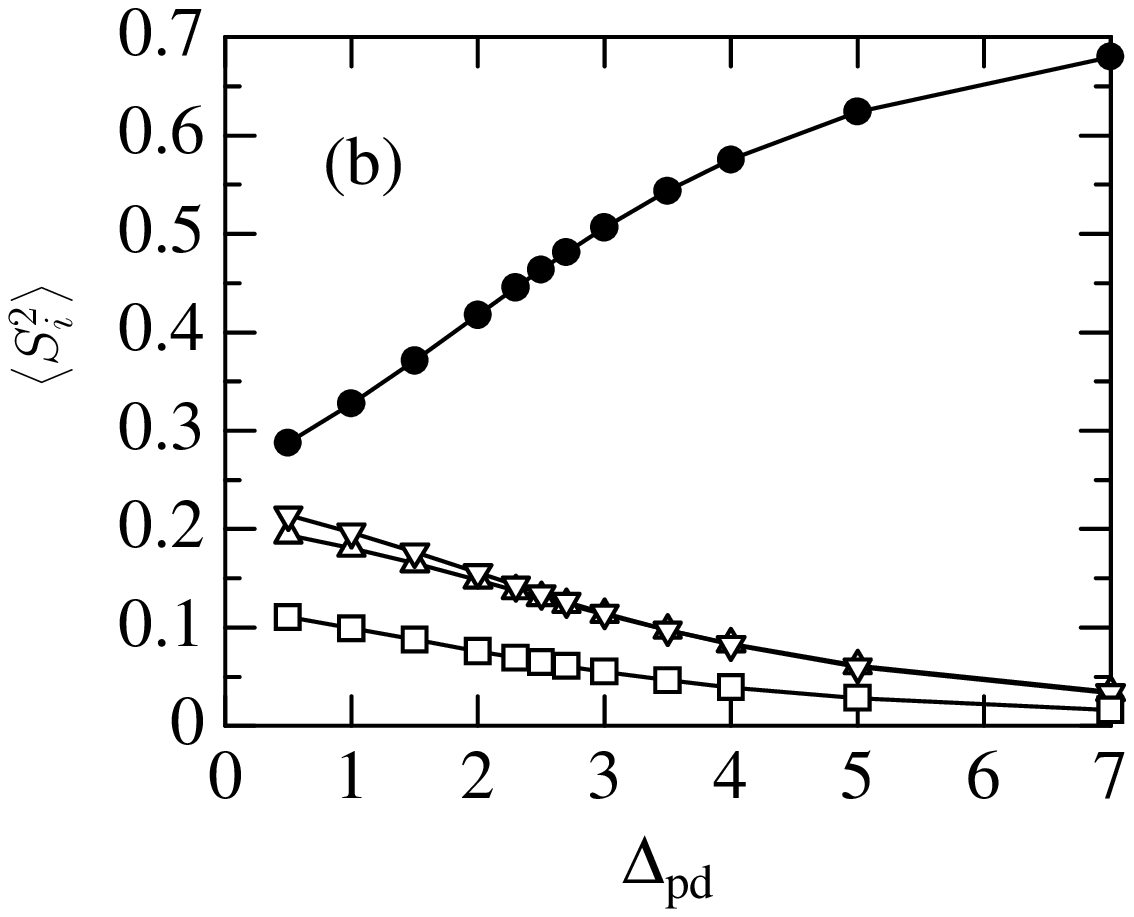}
\caption{\label{fig3} Local spin moment $\langle {\bm S}^2_i\rangle$
on the various types of sites
of an undoped ladder with $8\times 2$ Cu atoms:
Cu sites (solid circles), rung O($p_y$) sites (up-triangles), leg 
O($p_x$) sites (down-triangles),
and outer O sites (squares) versus (a) $t_{pp}$ for $\Delta_{pd}=2.7$
and (b) $\Delta_{pd}$ for $t_{pp}=0.5$.}
\end{figure}

The average value of the square of the
spin moment
\begin{equation}
\left\langle {\bm S}^2_i\right\rangle = \frac{1}{4}
\left\langle\left(\Psi^\dagger_{i\alpha} 
{\bm{\sigma}}_{\alpha\beta}
\Psi_{i\beta}\right)^2\right\rangle 
\end{equation}
on various sites is shown in Fig.~\ref{fig3}
for the same parameters. Here
$\Psi^\dagger_{i\alpha} = d^\dagger_{i\alpha}$ or
$p^\dagger_{i\alpha}$ depending upon the site, 
${\bm{\sigma}}$ is the usual Pauli spin matrix, and the indices
$\alpha$ and $\beta$ are summed over.   For
$\Delta_{pd} = 3$ and $t_{pp} = 0.5$, the average spin moment on
the Cu site is 0.5. Dividing this by the average hole occupation
for the Cu gives a squared spin moment of order 0.7 compared with
3/4 for a Heisenberg spin $s=1/2$. With $U_d=8$, the spin moment is well
formed on the Cu site and $\langle {\bm S}^2_{\text{Cu}}\rangle$ simply 
tracks
$\langle n_{\text{Cu}}\rangle$, the probability of having a hole on a 
Cu site.

\begin{figure}
\includegraphics[width=6cm]{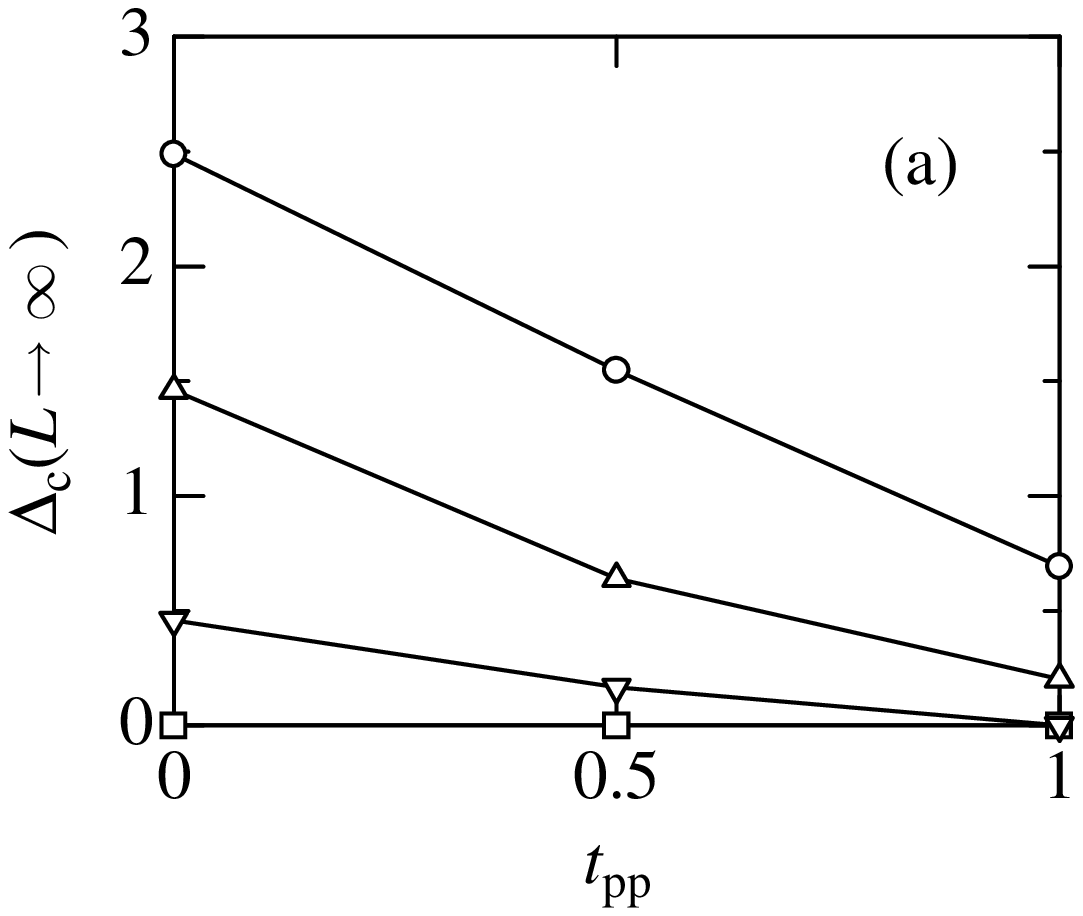}
\includegraphics[width=6cm]{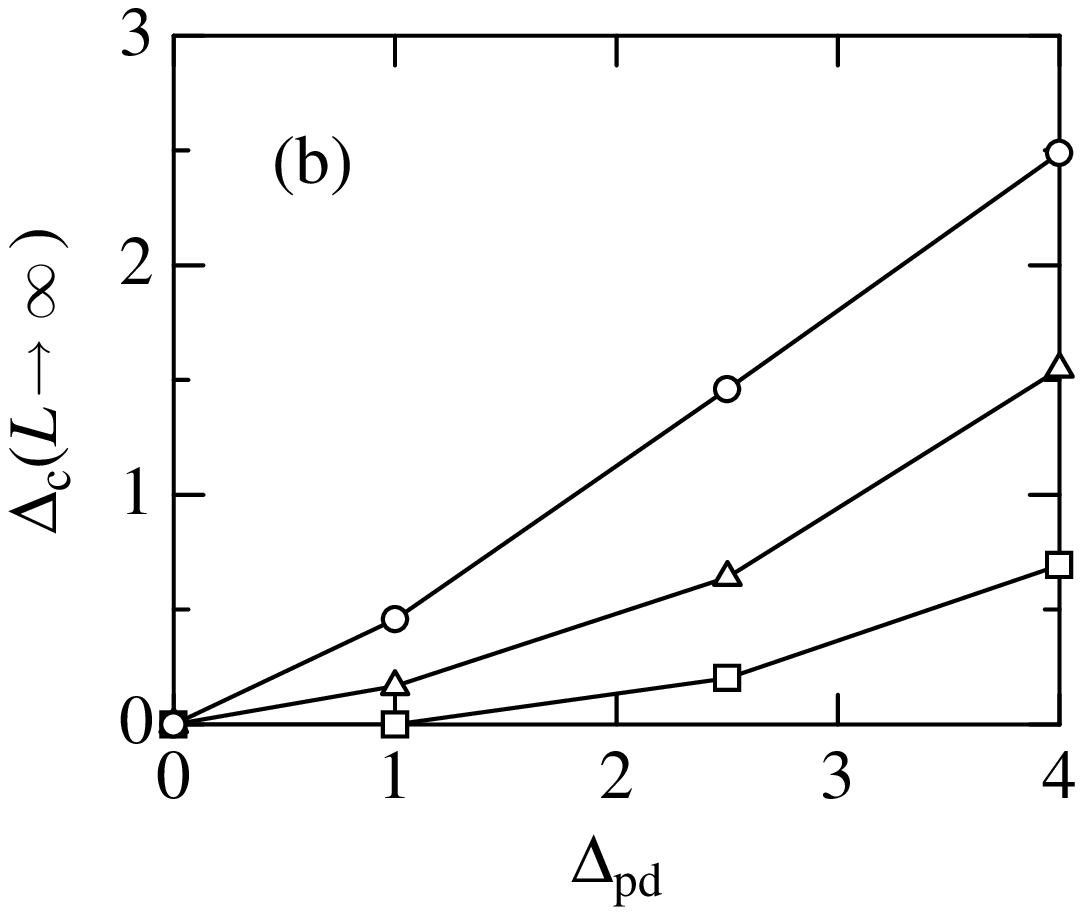}
\caption{\label{fig4} 
Charge gap of undoped ladders extrapolated to an infinite ladder 
length $L \rightarrow \infty$.
(a) As a function of $t_{pp}$ for $\Delta_{pd} = 0$ (squares), 
$\Delta_{pd} = 1$ (down triangles), $\Delta_{pd} = 2.5$ (up triangles),
and $\Delta_{pd} = 4$ (circles).
(b) As a function of $\Delta_{pd}$ for $t_{pp} = 0$ (circles), 
$t_{pp} = 0.5$ (triangles), and $t_{pp} = 1$ (squares).
}
\end{figure}

The charge gap of an undoped ladder with $L\times2$
Cu sites is given by
\begin{equation}
\Delta_c (L) = E_0 (+2,L) + E_0(-2,L) - 2E_0 (0,L)\ ,
\label{chargegap}
\end{equation}
where $E_0(N,L)$ denotes the ground state energy of 
a ladder of length $L$ with $N+2L$ holes. 
Extrapolating the charge gap 
from numerical data for up to $L=16$, we find the results
for $\Delta_c(L\to\infty)$
plotted in Fig.~\ref{fig4}. Here Fig.~\ref{fig4}(a) shows
$\Delta_c(L\to\infty)$ versus $t_{pp}$ for various
values of $\Delta_{pd}$ and Fig.~\ref{fig4}(b) shows $\Delta_c(L\to
\infty)$ versus $\Delta_{pd}$ for various values of
$t_{pp}$. From these results, we see that in order
to have a charge gap, we need $\Delta_{pd}\agt 2t_{pp}$.  
Experimentally~\cite{Pop00,Goz01}, $\Delta_c$ is of order
1.4 to 2 eV. 
If we take $t_{pp} \sim t_{pd}/2$ 
then $\Delta_{pd}\approx 3$ 
gives a realistic 
charge gap with $t_{pd}\approx 1.6$eV.~\cite{ALJP95}

\begin{figure}[t]
\includegraphics[width=6cm]{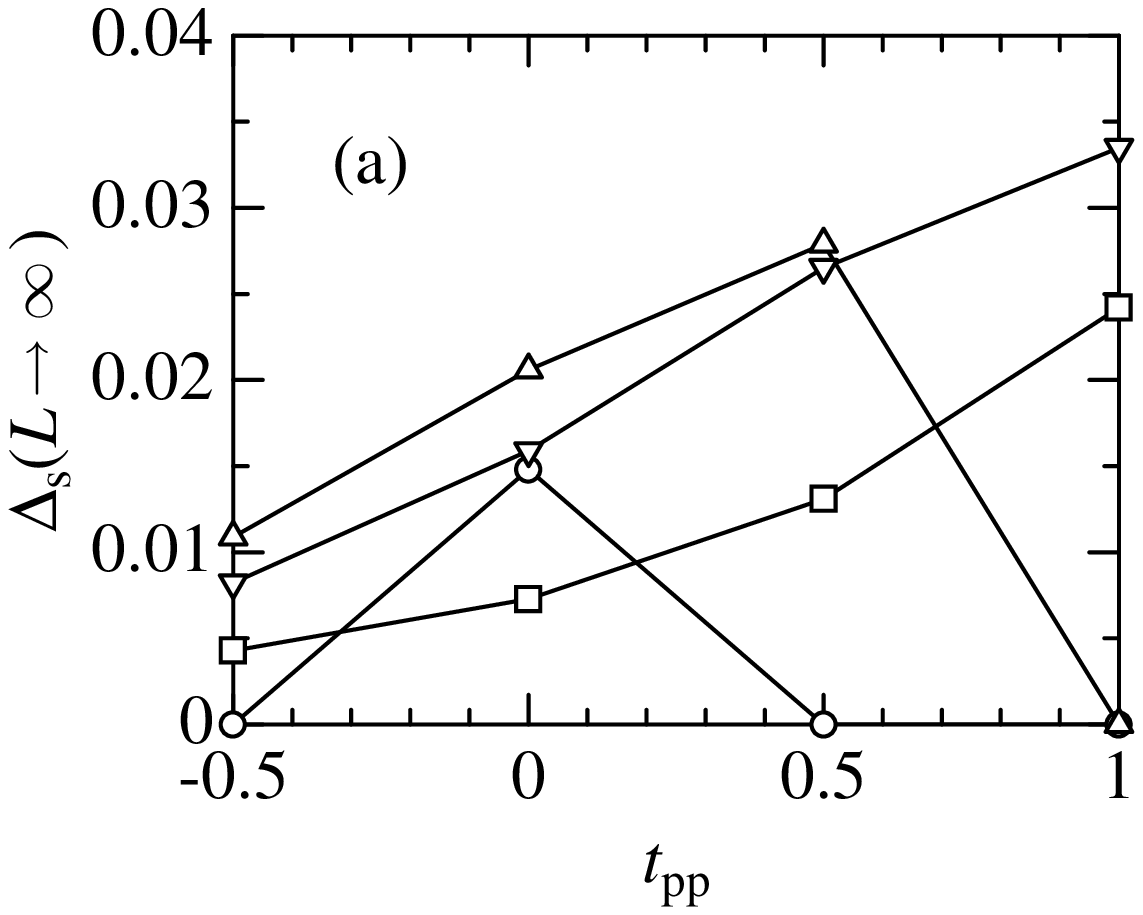}
\includegraphics[width=6cm]{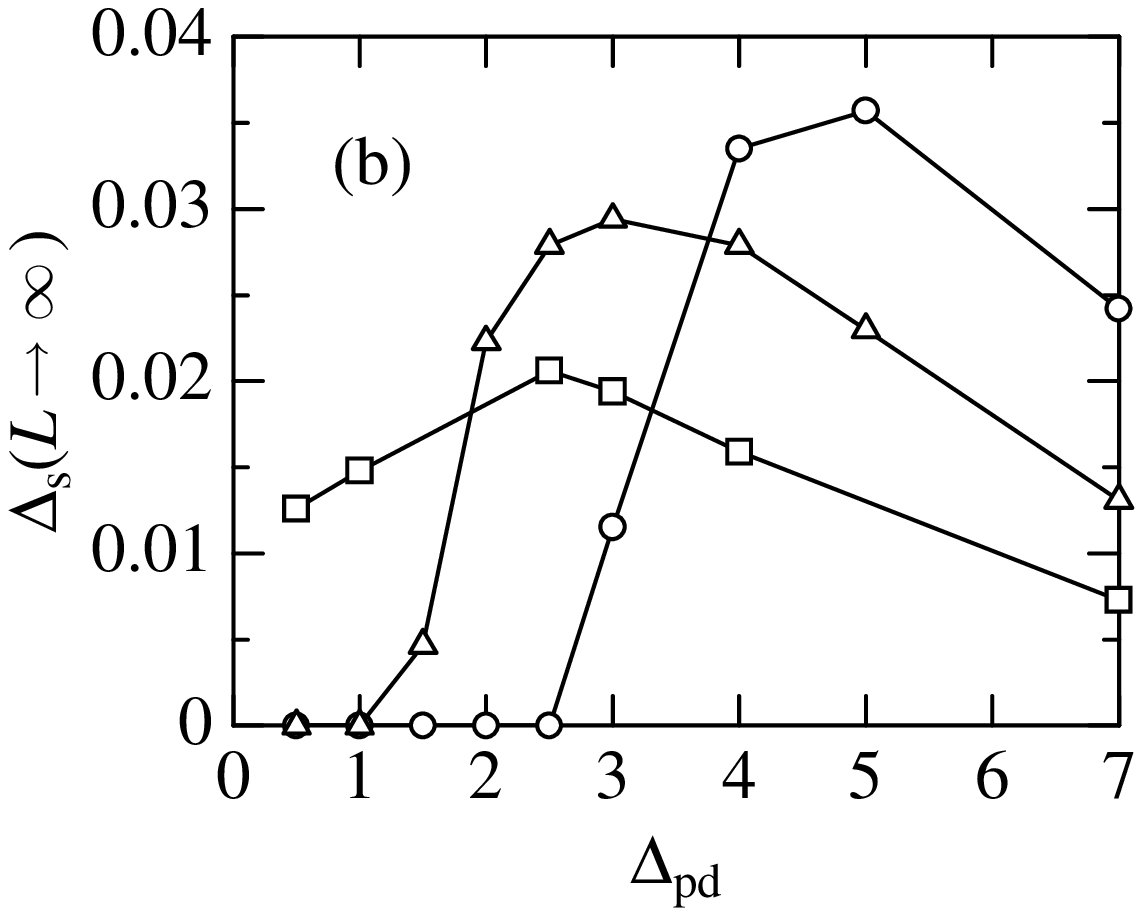}
\caption{\label{fig5} 
Spin gap of undoped ladders extrapolated to an infinite ladder length
$L \rightarrow \infty$.
(a) As a function of $t_{pp}$ for $\Delta_{pd}=1$ (circles),
$\Delta_{pd}=2.5$ (up triangles), $\Delta_{pd}=4$ (down triangles),
and $\Delta_{pd}=7$ (squares).
(b) As a function of $\Delta_{pd}$ for $t_{pp}=0$ (squares),
$t_{pp}=0.5$ (up triangles), and $t_{pp}=1$ (circles).}
\end{figure}

The lowest triplet spin excitation is gapped and localized
on the Cu sites of the ladder (i.e., the spin density
remains zero on the O sites in the triplet state).
The spin gap is given by
\begin{equation}
\Delta_s (N,L) = 
E_0\left (\frac{N}{2}+1, \frac{N}{2}-1, L \right )- 
E_0\left (\frac{N}{2},\frac{N}{2}, L \right) ,
\label{spingap}
\end{equation}
where $E_0(N_{\uparrow},N_{\downarrow}, L)$ is the ground state
energy of an $L\times2$ ladder with $N_{\sigma}+L$ holes of 
spin $\sigma=\uparrow,\downarrow$.  
We first discuss the spin gap of the undoped ladder 
($N=0$). 
In this case we extrapolate the spin gap to 
an infinite ladder length using data for up to $L=16$.
The dependence of the spin gap $\Delta_s(L \rightarrow \infty)$
on $t_{pp}$ and $\Delta_{pd}$ is shown in 
Figs.~\ref{fig5}(a) and \ref{fig5}(b),
respectively.  As for the charge gap, there is a spin gap  only
if $\Delta_{pd} \agt 2 t_{pp}$.
For large $\Delta_{pd}$ the spin gap decreases due to a decrease in 
the effective exchange coupling as observed
in our previous work.~\cite{JSW98}
Thus the spin gap goes through a maximum at a finite
value of $\Delta_{pd} >  2t_{pd}$.  
Experimentally,~\cite{Dag99,ALJP95} $\Delta_s\sim 0.03 -
0.05$eV so that with $t_{pd}\approx 1.6$eV,
we need
$\Delta_s/t_{pd}\simeq 0.03$.  This confirms that
$t_{pp}\simeq 0.5$ and $\Delta_{pd}\simeq 3$
represent appropriate choices, consistent with electronic band
structure calculations.~\cite{ALJP95}   

For a two-leg Heisenberg ladder, the spin gap is equal
to approximately half of the exchange interaction.
Thus from Fig.~\ref{fig5}(a) we see that provided 
$\Delta_{pd} \agt 2 t_{pp}$, the effective exchange interaction between
Cu spins increases with $t_{pp}$.
For example for $\Delta_{pd} =2.5$, the spin gap and hence the 
effective Cu-Cu exchange increase by 40 \% as $t_{pp}$ goes from 0
to 0.5. 
This is due to the additional exchange path through the O ions.
This crucial role of the O-O hopping $t_{pp}$ in giving a large 
effective Cu-Cu exchange
has been emphasized by Eskes and Jefferson.~\cite{EJ93}

The nearest-neighbor Cu spin-spin correlations 
$\langle {\bm S}_i \cdot {\bm S}_j \rangle$ are antiferromagnetic 
in the undoped ladder. For a fixed value of $t_{pp}$, 
$-\langle {\bm S}_i \cdot {\bm S}_j \rangle$ increases with 
$\Delta_{pd}$ and tends to the value obtained for
a two-leg 
Heisenberg ladder when $\Delta_{pd} \gg t_{pp}$.
This simply reflects the fact that increasing
$\Delta_{pd}$ causes the holes to become more
localized on the Cu sites, as shown by the increase
of the Cu spin moments $\langle {\bm S}_i^2\rangle$
in Fig.~\ref{fig3}(b). For the same reason, 
for a fixed value of $\Delta_{pd}$, 
$-\langle {\bm S}_i \cdot {\bm S}_j \rangle$
decreases with increasing $t_{pp}$ and vanishes for 
$t_{pp} \gg \Delta_{pd}$.
This decrease is due to the reduction of the local magnetic
moments $\langle {\bm S}_i^2\rangle$ on the Cu
sites as the holes spread further onto the O
sites, as seen in Fig.~\ref{fig3}(a).

The low energy properties of the spin degrees of
freedom of the undoped $d$-$p$ Hubbard model~(\ref{hamiltonian})
can be mapped
onto an anisotropic Heisenberg ladder.
\begin{equation}
H=\sum_{\langle ij\rangle} J_{ij}\ {\bm S}_i \cdot
{\bm S}_j\ .
\label{HeisHam}
\end{equation}
Here $\langle ij\rangle$ represent near-neighbor
bonds along the legs and rungs of a 2-leg ladder
with $J_{ij}=J_\perp$ for the rungs and
$J_\parallel$ for the legs. To estimate the
parameters $J_\perp$ and $J_\parallel$, we
consider the low-energy states of the $d$-$p$
model for a $\text{Cu}_{2L} \text{O}_{5L}$ cluster ($L\times 2$
Cu ladder). Then by making a one-to-one
correspondence of the states of this system to the
$L\times 2$ Heisenberg ladder,
we derive the
$J_\perp$ and $J_\parallel$ parameter of the
anisotropic Heisenberg Hamiltonian.  These
processes are performed for ladders of length
$L=8$, 12, and 16 with the open boundary
condition.  We confirm that the low-lying spin
states of the $d$-$p$ and Heisenberg clusters have
the same quantum numbers and characters for all
the three values of $L$.  To make the mapping systematically
with various system sizes, we extract the two
lowest triplet excitation energies, $E_{t1}=E_0 (S_{\rm
tot}=1) - E_0(S_{\rm tot}=0)$ and
$E_{t2}=E_1(S_{\rm tot}=1) - E_0(S=0)$. Here,
$E_n(S=S_{\rm tot})$ is the energy of the $n^{\rm
th}$
excited state with total spin $S$.  Then we
determine $J_\perp$ and $J_\parallel$ by fitting
the excitation energies $E_{t1}$ and $E_{t2}$ with
results obtained from $d$-$p$
clusters with the same $L$. In this way, with
$\Delta_{pd} = 3$ and $t_{pp}=0.5$,  we find
that $J_\perp=0.06t_{pd} (\sim 96$meV) and
$J_\parallel=0.08t_{pd} (\sim$ 128meV), both of
which are in a reasonable range as compared to
experimental and theoretical
results.~\cite{Pop00,Goz01,Gra99} 

\begin{figure}
\includegraphics[width=6cm]{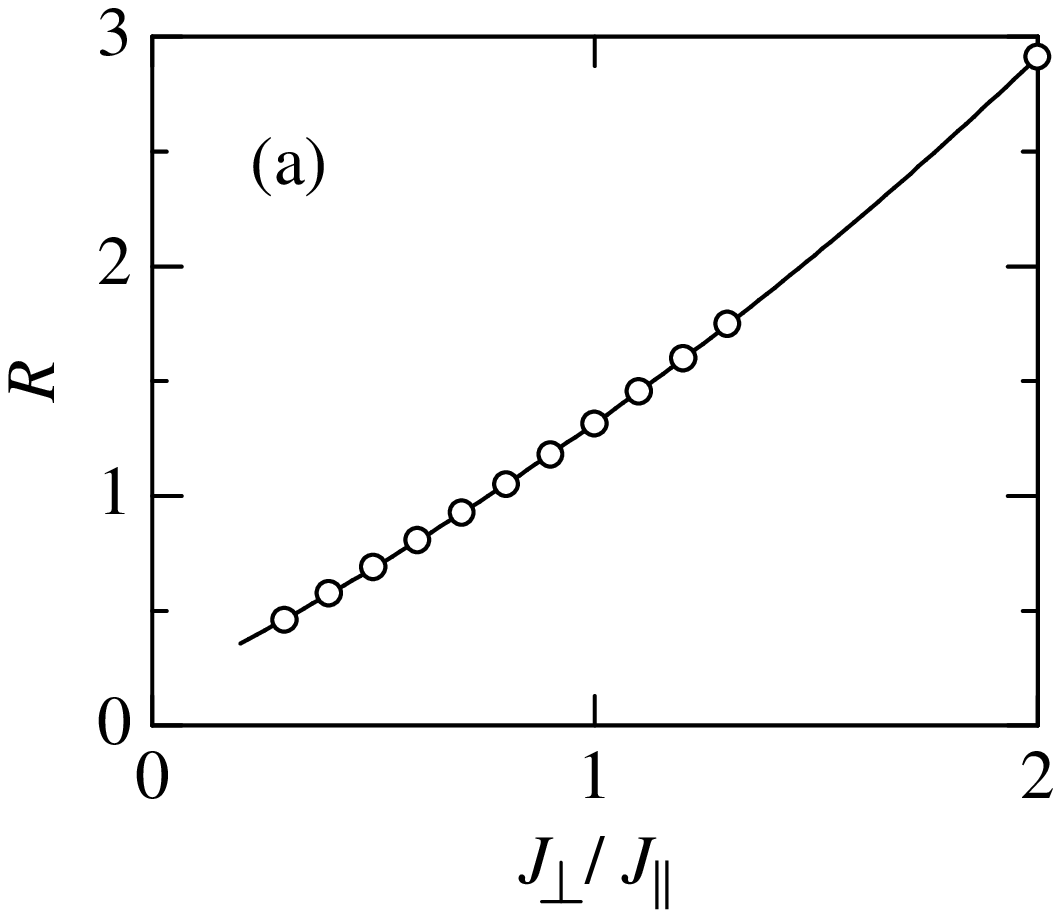}
\includegraphics[width=6cm]{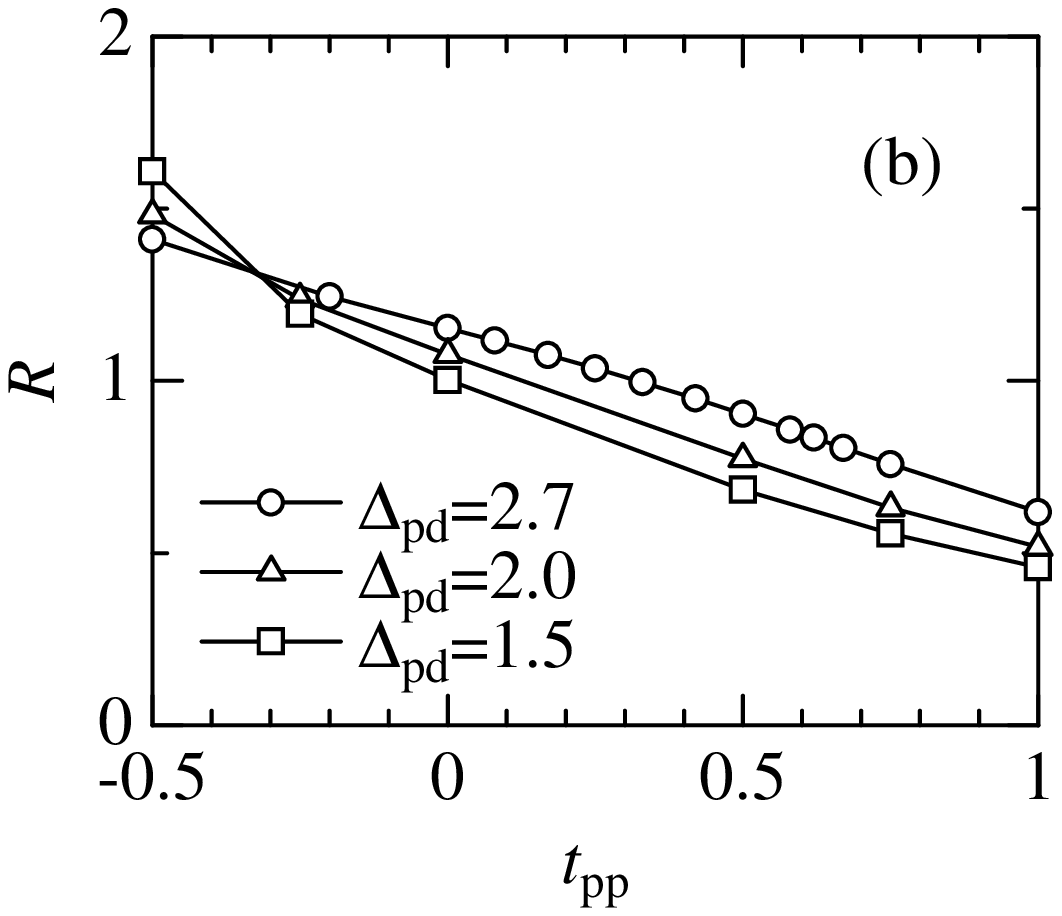}
\includegraphics[width=6cm]{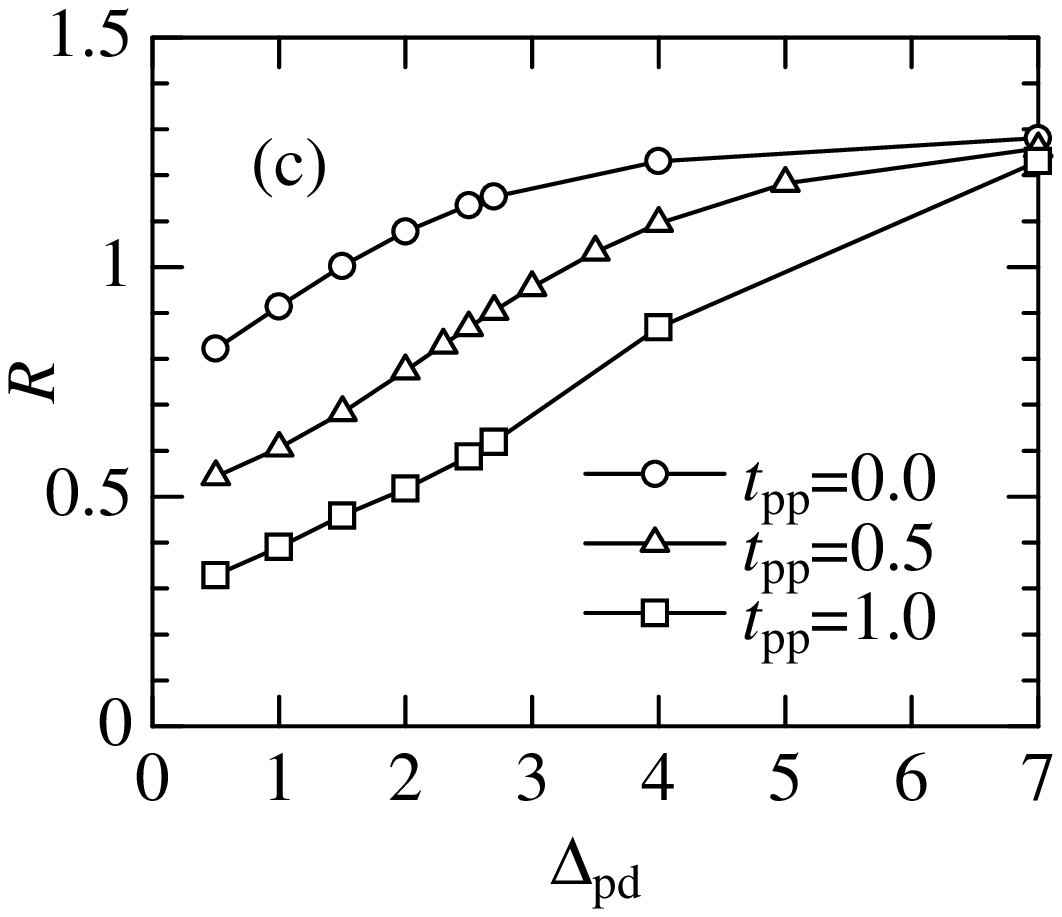}
\caption{\label{fig6} 
Ratio $R$ between nearest-neighbor spin correlations on rungs and legs.
(a) For a two-leg Heisenberg ladder with 100 rungs as a function
of $J_\perp/J_\Vert$.
For an undoped CuO ladder with $8\times2$ Cu sites (b) versus
$t_{pp}$ and (c) versus $\Delta_{pd}$. }
\end{figure}

It is also interesting to examine the effect
of $t_{pp}$ and $\Delta_{pd}$ on the effective
exchange coupling anisotropy.  From the nearest-neighbor Cu
spin-spin correlation $\langle {\bm S}_i \cdot {\bm S}_j \rangle$
we define
\begin{equation}
R=\frac{\left\langle {\bm S}_i \cdot {\bm S}_j
\right\rangle_{\text{rung}}} {\left\langle
{\bm S}_i \cdot {\bm S}_j 
\right\rangle_{\text{leg}}}, 
\label{spinratio}
\end{equation}
where we use an average of $\langle {\bm S}_i \cdot {\bm S}_j 
\rangle$ over the ladder.
For a Heisenberg ladder~(\ref{HeisHam}) with an exchange coupling
$J_\perp$ on the rungs and $J_\Vert$ on the legs, $R$
versus $J_\perp/J_\Vert$ is shown in Fig.~\ref{fig6}(a).  
Various experimental and theoretical
estimates~\cite{Pop00,Goz01,Gra99}
suggest that values of $J_\perp/J_\Vert$ ranging from 0.5 to
1.0 are appropriate for two-leg CuO ladders.  
Figure~\ref{fig6}(a) shows that this corresponds to 
values of $R$ varying from 0.6 to 1.2.  
The ratio $R$ obtained for our CuO ladder are shown 
versus $t_{pp}$ and $\Delta_{pd}$ in 
Figs.~\ref{fig6}(b) and \ref{fig6}(c), respectively. 
We see that the exchange coupling anisotropy is
again consistent with parameter values
$\Delta_{pd} \approx 3$ and
$t_{pp} \approx 0.5$. 
In the absence of the outer O sites, 
the anisotropy 
ratio $R$ becomes 
significantly larger than the values shown in Figs.~\ref{fig6}(b) and 
\ref{fig6}(c) for $ 0 < t_{pp} \leq 1$. In the charge transfer
insulating phase, one finds an effective exchange coupling ratio
$J_\perp/J_\Vert > 1$ for realistic parameters, in disagreement
with most experimental and theoretical estimates.  
Thus, it appears that it is important to include the outer O sites 
in order to obtain a qualitatively correct 
description of two-leg CuO ladders.

The dependence of the charge and spin
magnitude distributions, the 
charge and spin gaps, and the
spin-spin correlations on the parameters $t_{pp}$ and $\Delta_{pd}$
can be understood qualitatively.
For $t_{pp} \agt \Delta_{pd}/2$, the holes are 
delocalized on all Cu and O sites
and the undoped CuO ladder is in a metallic phase.
For $\Delta_{pd} \agt 2t_{pp}$, the holes tend to
be localized on the Cu sites
and the undoped system is a charge-transfer
insulator. 
Low-energy spin excitations involve the holes on the Cu sites
only.
Thus, the spin degrees of freedom localized on the Cu sites
lead to an effective two-leg spin ladder.
Here, as noted,
we are working in units where $t_{pd}=1$, $U_d=8$, and $U_p=3$.
We have seen that the magnitudes of the charge and spin
gaps as well as the site charge densities, spin moments, and 
effective exchange couplings are consistent with $\Delta_{pd}
=3$ and $t_{pp} = 0.5$. We will now examine what happens when
such a CuO ladder is doped with holes and electrons.  

\section{The Doped Ladder \label{sec:doped}}

\begin{figure}
\includegraphics[width=6cm]{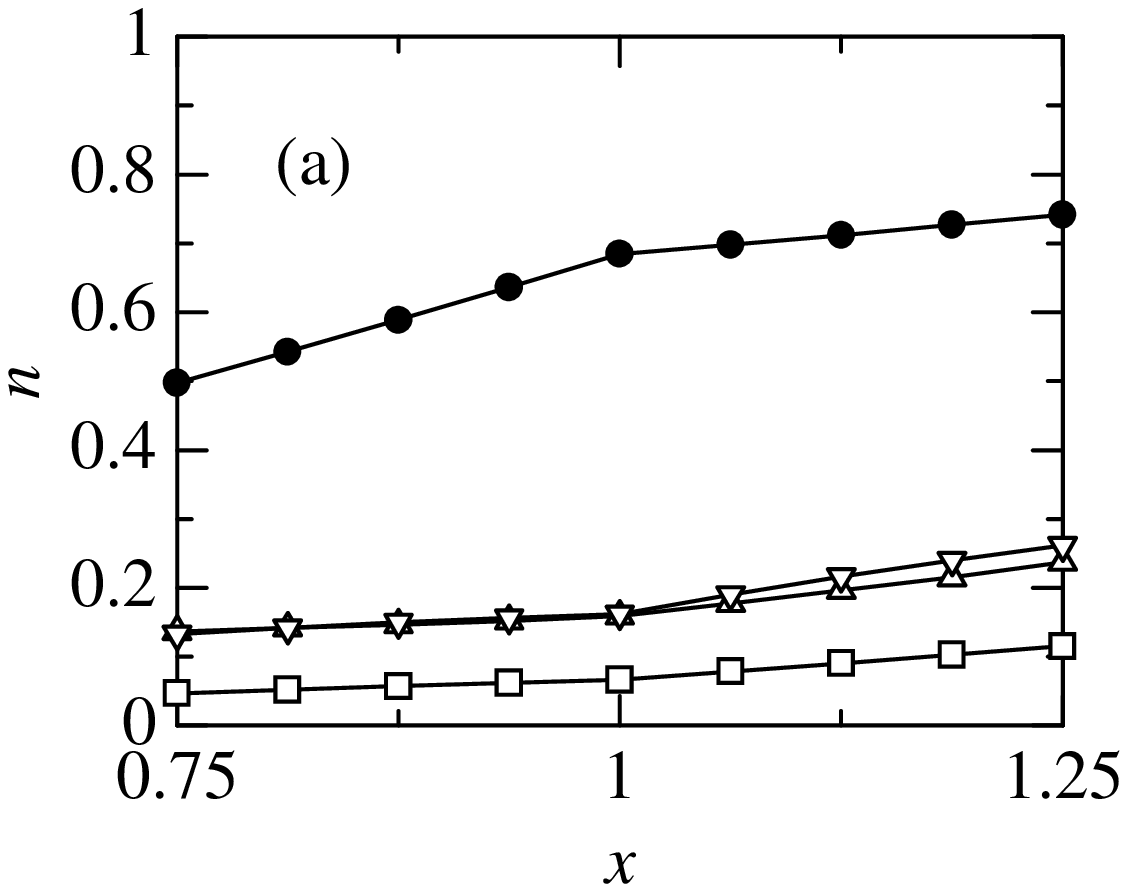}
\includegraphics[width=6cm]{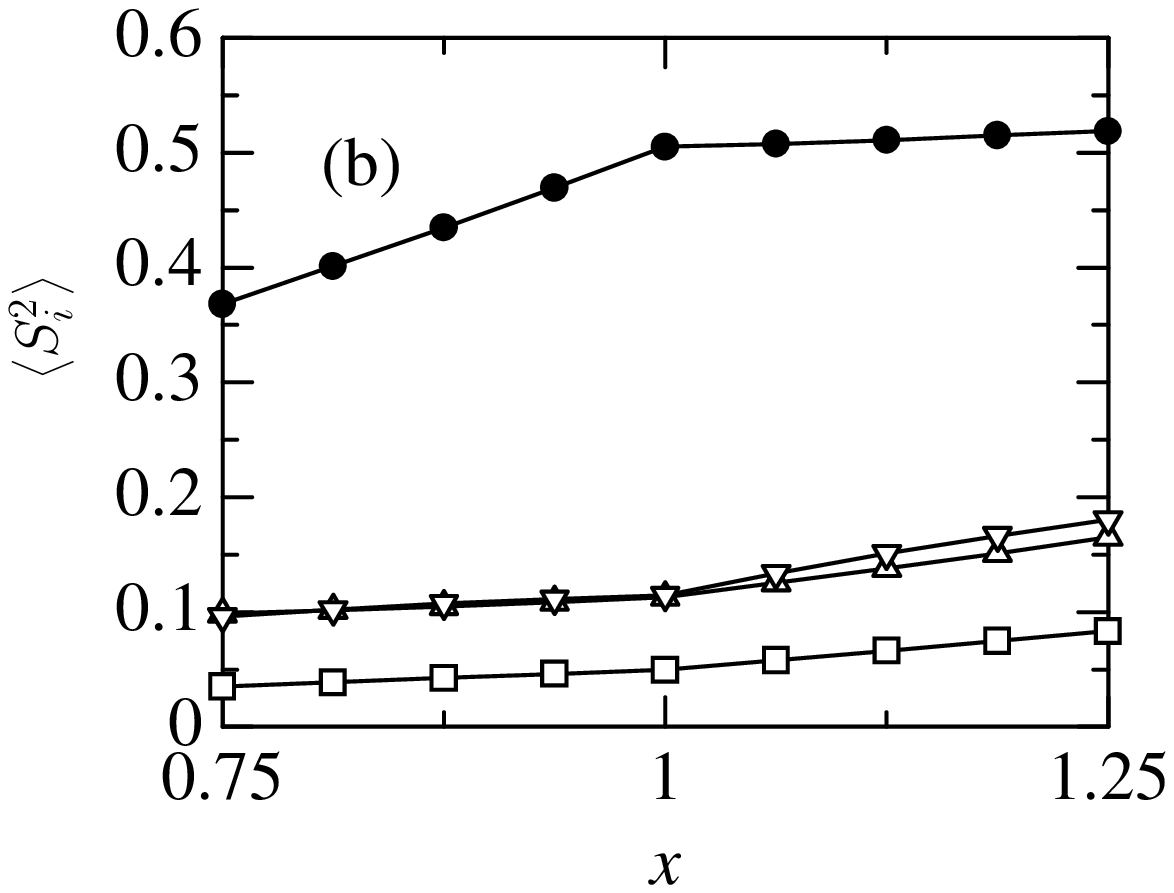}
\caption{\label{fig7} (a) The hole density $n$ and (b) the square of the
spin moment $\langle {\bm S}^2_i\rangle$ on the various types of sites
of a ladder with $16\times 2$ Cu atoms as a function of filling $x$ :
Cu sites (solid circles), rung O($p_y$) sites (up triangles), leg 
O($p_x$) sites (down triangles), and outer O sites (squares).}
\end{figure}

Turning to the doped situation, we investigate the properties
of the model~(\ref{hamiltonian}) with $\Delta_{pd}=3$,
$U_d=8$, $U_p=3$, and $t_{pp}=0.5$ for various hole concentrations
(per Cu atom) $x=1+N/(2L)$, where $N$ is the number of doped
holes ($N>0, x>1$) or doped electrons ($N < 0, x <1$) in a ladder
with $L \times 2$ Cu atoms.
Results for the site occupations versus the doping are shown in 
Fig.~\ref{fig7}(a).
Here, we see that the slope of the hole
occupation on the Cu site versus $x$ changes at $x=1$.
For electron doping, the electrons are more likely to go
on the Cu site than are the holes when the system is
hole doped.  This is, of course, what one would expect
for a charge-transfer insulator.  This is also
consistent with Monte Carlo results for the 3-band
Hubbard model~\cite{Sca91,DMH92}
 as well as our previous DMRG results for
the CuO ladder.~\cite{JSW98}  
The variation of the local squared spin moments on the different sites
versus doping are shown in Fig.~\ref{fig7}(b). 
As discussed previously, in the charge
transfer regime, where $U_d$ is large compared to $\Delta_{pd}$ and
$\Delta_{pd}$ is larger than several times $t_{pp}$, the spin moments 
are mainly on the Cu sites.  
When the system is electron doped, the electrons
go onto the Cu sites eliminating their moments.  
Thus, the
average moment on the Cu sites depends upon $x$ in the same way as the 
Cu site occupation $\langle n_{\text{Cu}}\rangle$, and 
$\langle {\bm S}^2_{\text{Cu}} \rangle
/\langle n_{\text{Cu}}\rangle$ remains essentially constant, of order
0.7. 
The change in the average moment shown in Fig.~\ref{fig7}(b) 
simply reflects the fact that for the electron-doped case, 
spin moments at individual Cu sites are removed when electrons are 
added to those sites. Likewise, for the hole-doped situation 
the average spin moment on the Cu sites increases only slightly as $x$
increases and a few more Cu sites have holes. 

\begin{figure}
\includegraphics[width=6cm]{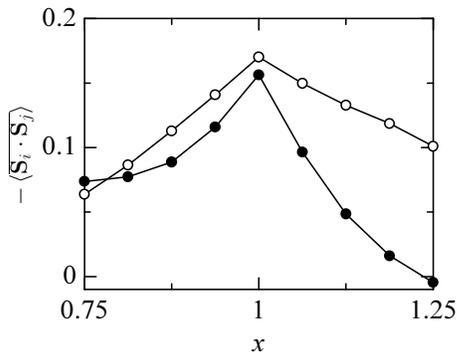}
\caption{\label{fig8} 
The average nearest-neighbor Cu spin-spin correlation function $\langle
{\bm S}_i \cdot {\bm S}_j\rangle$ on the rungs (solid circles)
and legs (open circles) versus $x$ calculated on ladders with
$16\times 2$ Cu sites.}
\end{figure}

In Fig.~\ref{fig8}
we have plotted the average value of the nearest-neighbor Cu
spin-spin correlations $\langle {\bm S}_i \cdot {\bm S}_j\rangle$
on the rungs and legs versus doping. As
electrons are doped onto the ladder and spin moments are removed from the
Cu sites, the magnitude of the $\langle {\bm S}_i \cdot {\bm S}_j\rangle$
correlations decrease on both the rungs and the legs.  
Similar results are observed for the nearest-neighbor spin-spin 
correlations in the $t-J$ model on a two-leg ladder
as the hole concentration increases.

However, as holes are added to the CuO Hubbard model (\ref{hamiltonian})
and $x$ increases, the rung spin-spin correlations decrease more
rapidly than those along the legs. As we have seen, 
the added holes tend to 
go onto the O sites which frustrates~\cite{Aha88}
 or screens the exchange coupling between
the Cu spins. A rough estimate of this can be seen by considering a
$\text{Cu}_4 \text{O}_{12}$ cluster with one hole or electron added.  
Calculating the
low-lying states of the cluster and comparing the excitation energies with
a 4-site $t$-$J_{\text{eff}}$ system we find that $J_{\text{eff}}/t_{pd}$ 
is 50\%
smaller for the one hole $(x=1.25)$ case relative to its value for the one
electron $(x=0.75)$ case.
Thus, in the hole-doped system, the effective
exchange interaction between the spin moments on the Cu sites is weakened.
Because this effective exchange is initially larger along  the legs than
across the rungs, the rung correlations decrease more rapidly than those
along the legs. 
We note that the anisotropy in the reduction of the spin-spin 
correlations
as a function of doping is much stronger than in a $t-J$ model
on a two-leg ladder,
even if one uses a strongly anisotropic exchange coupling 
$J_{\perp} \neq J_{\parallel}$.

The average values of the nearest-neighbor Cu-O spin-spin correlations
increase rapidly with the number of holes in the ladder.
For hole doping the appearance of
strong nearest-neighbor Cu-O spin correlations
corresponds to the formation
of a Zhang-Rice singlet~\cite{ZR88} in an isotropic two-dimensional 
$\text{CuO}_2$ lattice.
In the two-leg CuO ladder, however, the strength of the Cu-O spin-spin
correlation depends on the type of the O site considered.
Rung Cu-O spin-spin correlations increase less rapidly than the other
ones with increasing hole concentration $x$. This confirms the higher
frustration of antiferromagnetic correlations in the rung direction
than in the leg direction when an additional hole is placed between
the two holes localized on nearest-neighbor Cu sites.
Correspondingly, for hole doping the spin-spin correlation between a Cu 
site and its outer O site increases faster than for the other 
nearest-neighbor sites because of the absence of frustration for
this type of O sites.

\begin{figure}
\includegraphics[width=7cm]{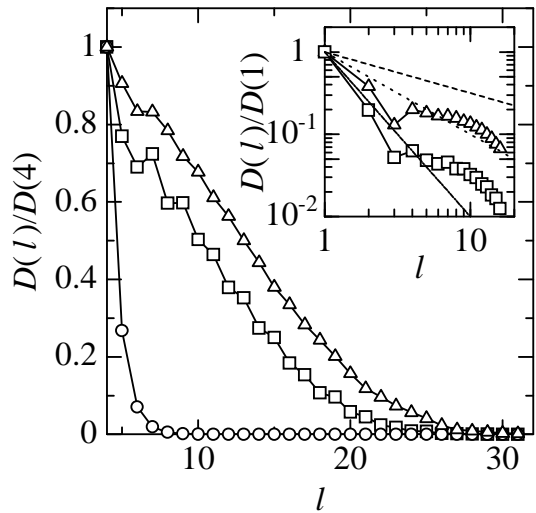}
\caption{\label{fig9} 
The Cu rung-rung pair field correlation function $D(\ell) = \langle
\Delta_{i+\ell} \Delta^\dagger_i\rangle$ versus $\ell$ for
a $32 \times 2$ Cu ladder. Here, we have 
normalized $D(\ell)$ to its value at $\ell=4$.
Circles, squares, and triangles correspond to an undoped ladder,
a ladder doped with two holes, and with two electrons, respectively.
Inset: Same data (normalized to their $\ell=1$ value)
on a log-log scale. The lines have slope $-1/2$, $-1$ and $-2$. 
}
\end{figure}

Turning next to the pairing correlations, we have
calculated the Cu rung-rung pair field correlation function
\begin{equation}
D(\ell) = \left\langle \Delta_{i+\ell} \Delta^\dagger_i
\right\rangle
\label{njsfour}
\end{equation}
with
\begin{equation}
\Delta^\dagger_i = d^\dagger_{i1\uparrow}
d^\dagger_{i2\downarrow} - d^\dagger_{i1\downarrow}
d^\dagger_{i2\uparrow}\ .
\label{njsfive}
\end{equation}
Here, $d^\dagger_{i\lambda s}$ creates an electron of
spin $s$ on the $i^{\rm th}$ rung and the $\lambda=1$ or
2 leg of the ladder. The pair field has
``$d$-wave-like''
structure in the sense that the
rung-leg Cu-Cu pair field correlation
function, has a negative sign while the
rung-rung or leg-leg pairing correlations
are positive. 
We have also examined other pair field correlation functions
corresponding to pairs on diagonal Cu sites or pairs on O sites
and found that they are either qualitatively similar to the Cu 
rung-rung pair field correlations or decrease much faster. 
Some results for the rung-rung pair field correlation
function $D(\ell)$ versus $\ell$ 
are shown in Fig.~\ref{fig9} 
for a ladder containing $32\times 2$ Cu sites.
We have normalized $D(\ell)$ with
respect to its value at $\ell=4$. The pair field
correlations in the undoped system decay rapidly (exponentially) with
distance while, as shown in the inset, the pair correlations when two
electrons are added or removed exhibit an approximate power law decay.
(Qualitatively similar pair field correlation functions have been
calculated in a $t-J$ two-leg ladder with two doped holes.)
While the decay of the normalized rung-rung pair field correlations for
the 2-electron and 2-hole doped ladders are similar in the
CuO Hubbard model~(\ref{hamiltonian}), the size of the 
rung-rung Cu pairing correlations are a factor of order 4 larger for the
electron-doped case relative to the hole-doped one. This 
reflects the fact that for electron doping, the added carriers go
primarily onto the Cu sites while for hole doping they 
go primarily onto the O sites.
For higher doping ($0.75 \leq x \leq 1.25$) the Cu rung-rung
pair field correlations are very similar to those shown in 
Fig.~\ref{fig9} for doped holes or electrons.

To determine whether doped holes or electrons form bound pairs
we have first calculated the binding energy of two doped holes
and electrons.
The pair binding energy is defined as
\begin{equation}
\Delta_{pb}(L) = 2 E_0(\pm 1, L) - E_0(\pm 2,L) -E_0(0,L) ,
\label{bindingenergy}
\end{equation}
where $E_0(N,L)$ is the ground state energy of a ladder with 
$L \times 2$ Cu site and $N+2L$ holes.  
We could not extrapolate $\Delta_{pb}(L)$ to the 
$L \rightarrow \infty$ limit because finite-size effects are too 
irregular. However, we have found that 
$\Delta_{pb}(L\rightarrow \infty)$ is certainly positive
both for hole and electron doping.
Therefore, we expect two doped holes or electrons to build
a bound pair in the CuO Hubbard ladder for the parameters
we use here.
As in our previous work,~\cite{JSW98} the binding energy
of two doped holes is of the order of the spin gap of the undoped 
ladder. 
For instance, for a $16 \times 2$ Cu site ladder we have obtained 
$\Delta_{pb} \approx 0.02 t_{pd}$, to be compared with the spin gap
$\Delta_{s}(L\rightarrow \infty) = 0.03 t_{pd}$.
The binding energy of two electrons, $\Delta_{pb} \approx 0.04 t_{pd}$,
is also of the order
of the spin gap in the undoped ladder but twice as large as
the hole pair binding energy. 
In our previous investigation of the two-leg CuO ladder~\cite{JSW98}
we found a vanishing pair binding energy for electron
doping. 
To understand this change  
we have calculated  the pair binding energy as a function
of the various model parameters in a ladder
with $8 \times 2$ Cu sites. 
It appears that the difference between the present result 
and the previous one~\cite{JSW98}
 is due to the nearest-neighbor O hopping $t_{pp}$.
In fact, $t_{pp} > 0$ is necessary to obtain
a finite pair binding energy in the electron doped ladder.  
When the CuO model is reduced to a simple,
one-band Hubbard model, $t_{pp} > 0$ leads
to an effective next-near-neighbor
hopping $t^\prime$ which favors pairing when the system
is electron doped.~\cite{WS99}
In addition, as previously discussed, $t_{pp} > 0$ leads to an enhanced
Cu-Cu effective exchange.

\begin{figure}
\includegraphics[width=6cm]{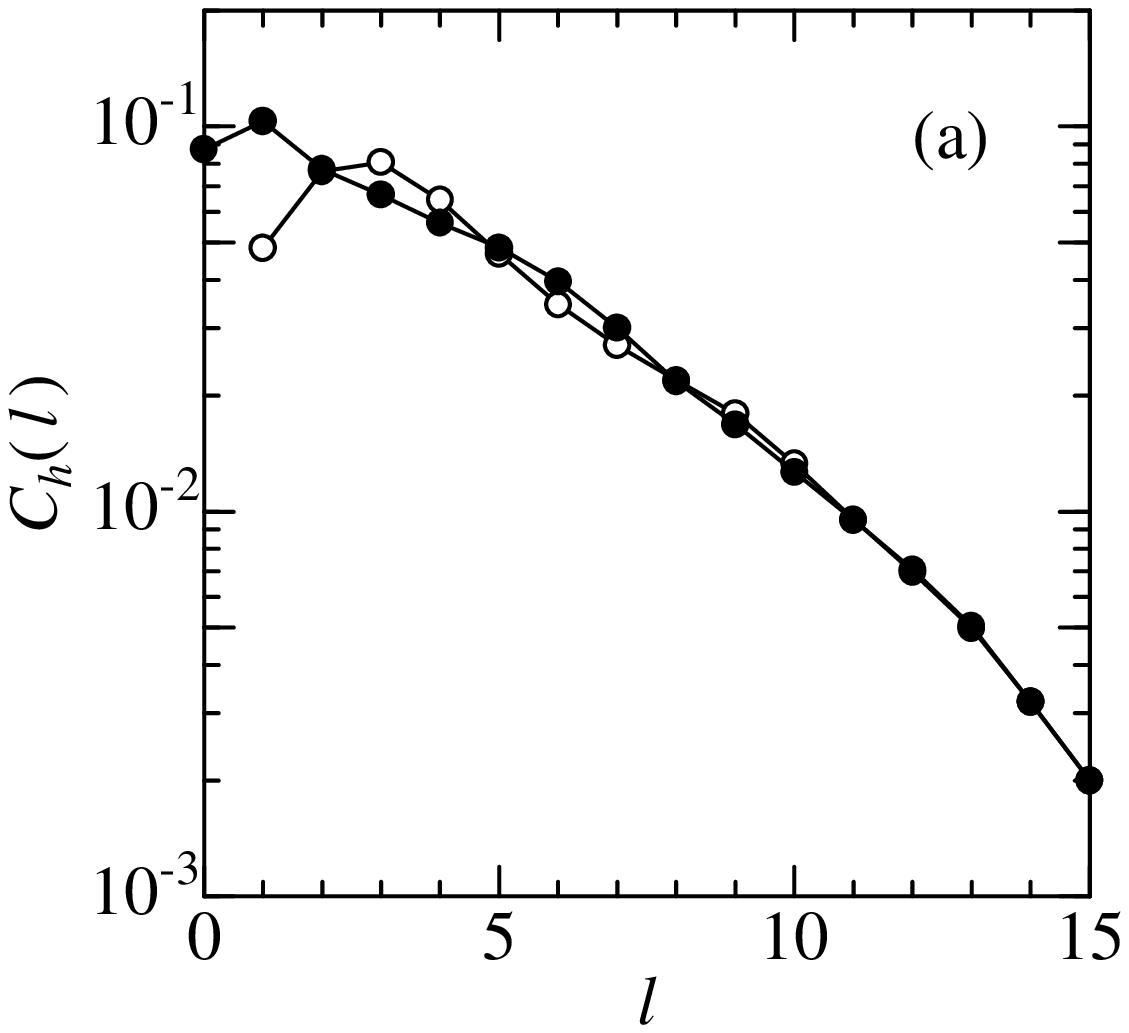}
\includegraphics[width=6cm]{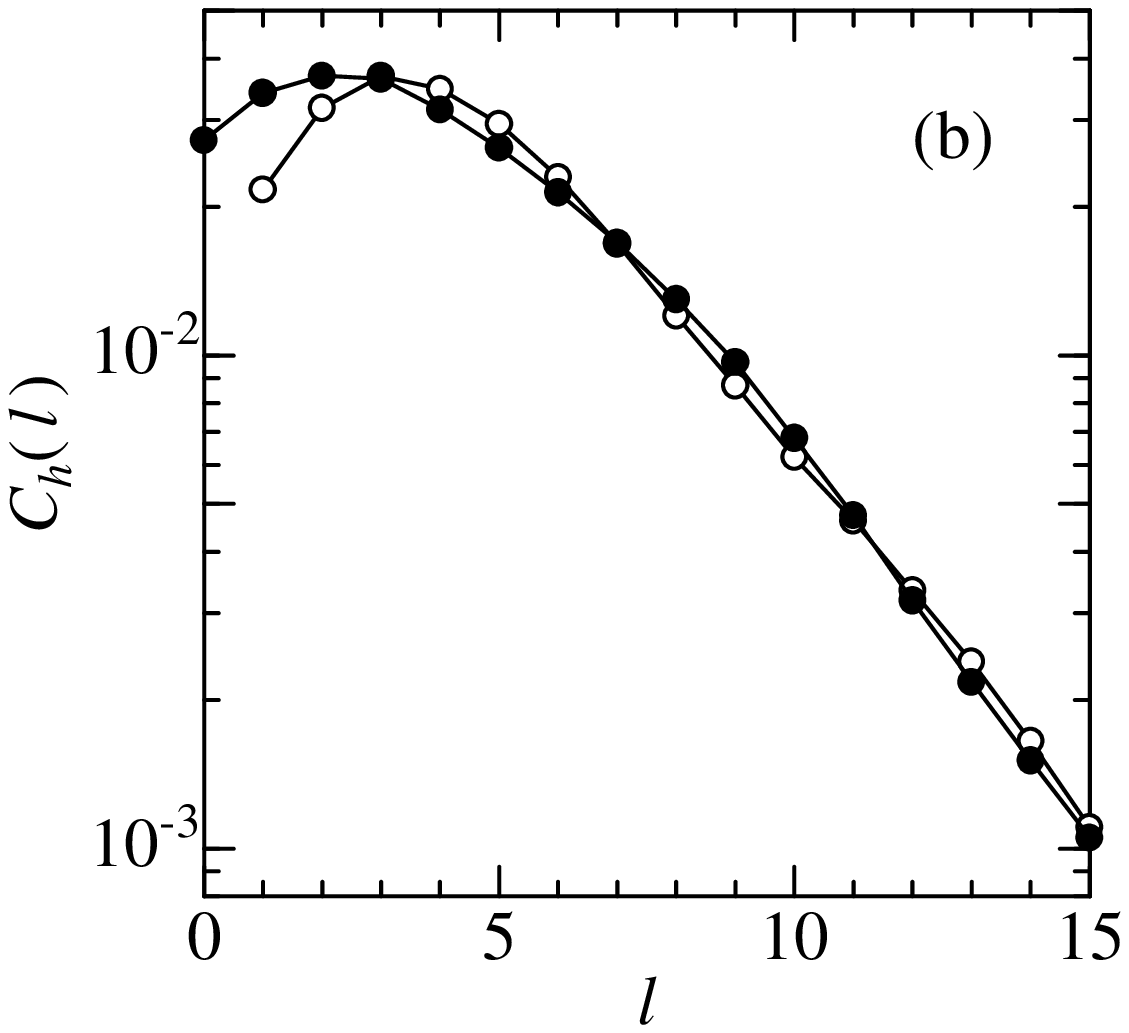}
\caption{\label{fig10}
The projected correlation functions $C_{e,h}(l)$
for (a) the Cu sites
of a two-electron doped ladder and (b) the $\text{O}_4$ orbitals of 
a two-hole doped ladder. 
The distance $l$ is measured along the ladder in units of the interval
between two Cu rungs. 
Open and solid circles correspond to correlations between doped 
charges on the same leg and on different legs, respectively.}
\end{figure}

In order to have a more detailed picture of the difference between
the electron- and hole-doped pairs, we have investigated their internal 
structure. 
In a $t-J$ model or a spin-fermion model, it is easy to measure
correlations between doped charges because there are no other
itinerant charges (i.e., there are no charge fluctuations in the
undoped system).
In the CuO Hubbard model~(\ref{hamiltonian}), however,
bare electron-electron and
hole-hole correlation functions give little information about the 
correlations between two doped charges because they are dominated by 
the quantum fluctuations of the other charges.
Therefore, we have to project out the ground state 
of~(\ref{hamiltonian})
onto a subspace with no charge fluctuation.
For that purpose we consider a perturbation expansion in $t_{pd}$ and
$t_{pp}$ of the Hamiltonian~(\ref{hamiltonian}) with
$U_d > \Delta_{pd} > 0$ and $x$ close to 1.
The ground state of the unperturbated Hamiltonian ($t_{pd}=t_{pp}=0$)
is degenerate. 
For electron doping the ground state has at most one hole on each
Cu site and no hole on any O site. 
The projection operator onto the ground-state subspace is
\begin{equation}
P_e = \prod_i (1-d^{\dag}_{i \uparrow} d_{i \uparrow} 
d^{\dag}_{i \downarrow} d_{i \downarrow}) \,  \prod_j 
(p_{j \uparrow} p^{\dag}_{j \uparrow} p_{j \downarrow} 
p^{\dag}_{j \downarrow}) \ .
\end{equation}
For hole doping there is exactly one hole on each Cu site
and no doubly occupied O site in the ground state of the 
unperturbated Hamiltonian.
The projection operator onto the ground-state subspace is
\begin{equation}
P_h = \prod_i (S^d_{i \uparrow} 
+ S^d_{i \downarrow})
\prod_j (1-p^{\dag}_{j \uparrow} p_{j \uparrow} 
p^{\dag}_{j \downarrow} p_{j \downarrow}) \ , 
\end{equation}
where $S^d_{i \sigma} = d^{\dag}_{i \sigma} d_{i \sigma}
d_{i, -\sigma} d^{\dag}_{i, -\sigma}$.
Using perturbation theory one could then derive effective 
Hamiltonians (a generalized $t-J$ model and a generalized spin-fermion 
model, respectively)
in the subspace defined by $P_e$ and $P_h$, which 
approximately describe the low-energy properties of the 
CuO model~(\ref{hamiltonian}) in the regime
$t_{pd}, t_{pp} \ll \Delta_{pd} < U_d$
for electron doping and hole doping, respectively.
Therefore, for the realistic parameters 
$\Delta_{pd} = 3 t_{pd} = 6 t_{pp}$, 
we calculate the ground state
$|\psi\rangle$ of (\ref{hamiltonian}) 
and then use the projected states $P_e |\psi \rangle$    
and $P_h |\psi \rangle$ 
to determine the correlations between doped carriers
(electrons or holes).

Figure~\ref{fig10}(a) shows the electron-electron correlation function 
\begin{equation}
C^n_e(m) = \frac{\langle \psi | P_e E_n E_{n+m} P_e| \psi  \rangle  }
{\langle \psi | P_e  E_n P_e | \psi \rangle } \ ,
\end{equation}
where $E_n = d_{n \uparrow}  d^{\dagger}_{n \uparrow}
d_{n \downarrow} d^{\dagger}_{n \downarrow}$,
calculated in a ladder with $32 \times2$ Cu atoms and
two doped electrons.
$C^n_e(m)$ measures the correlation between two added electrons
on the Cu sites.
Note that the most probable
arrangement for the two electrons is on diagonal nearest-neighbor
sites, which is similar to what is seen in t-J ladders. 
Results for the
hole-doped case are shown in Fig.~\ref{fig10}(b).
Here the hole-hole correlation function is given by 
\begin{equation}
C^n_h(m) = \frac{\langle \psi | P_h \sum_{\sigma} H_{n \sigma} 
H_{n+m, -\sigma} 
P_h | \psi  \rangle }
{\langle \psi | P_h  \sum_{\sigma} H_{n\sigma} P_h | \psi \rangle } \ ,
\end{equation}
where the operator 
\begin{equation}
H_{n \sigma} =  \frac{1}{4} \sum _{\langle n j \rangle}
p^{\dagger}_{j \sigma} p_{j \sigma} 
\end{equation}
measures the probability of finding
a hole in an $\text{O}_4$ orbital around the Cu site with 
index $n$
(the sum runs over the four
nearest-neighbor O sites of this Cu site).  
We average the density over four O sites around each Cu site
because a doped hole is locked in a singlet state with 
another hole on a Cu site as in a $\text{CuO}_4$ cluster 
(i.e., in a Zhang-Rice singlet~\cite{ZR88}).
This also facilitates the comparison with the electron-doped case and 
the $t-J$ two-leg ladder.
$C^n_h(m)$ measures the correlation between two added
holes in $\text{O}_4$ orbitals around different Cu sites.  
Figure~\ref{fig10}(b) shows that the most probable arrangement for the 
holes is on diagonal next-nearest-neighbor $\text{O}_4$ orbitals.

For larger distance $m$, both $C^n_e(m)$ and $C^n_h(m)$ decrease 
approximately as $\exp (-|m|/\xi)$ with correlation lengths 
(in units of the distance between two Cu rungs)  
$\xi_e \approx 3.3$ and $\xi_h \approx 2.9$, respectively. 
Note that these correlation lengths cannot be interpreted as the 
pair size because the exponential decay sets in only for
$m \agt \xi_{e,h}$.
The average distance between doped carriers is given by 
\begin{equation}
\overline{m}_{e,h} = \frac{\sum_m |m| C_{e,h}(m)}{\sum_m C_{e,h}(m)} \ ,
\end{equation}
which yields $\overline{m}_e \approx 4.5$ and 
$\overline{m}_h \approx 4.8$
for the electron and hole pair, respectively.

\begin{figure}
\includegraphics[width=7.2cm]{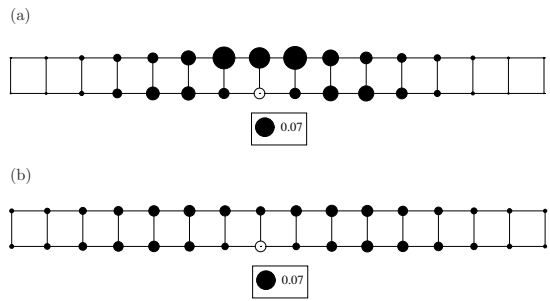}
\caption{\label{fig11}
A schematic view of the density-density correlations
showing the structure of a pair in (a) an electron-doped ladder
and (b) a hole-doped ladder.
Here, one carrier (doped
electron or hole) is located at the site $n$ marked by an open circle
(around the middle of the ladder). 
The radius of the solid circles is proportional to the probability 
$C^n_e(m)$ [$C^n_h(m)$] 
of finding the second doped electron (hole) on that site.
Only the central part of ladders with $32 \times 2$ Cu sites
is shown.
}
\end{figure}

A more intuitive picture of the difference in the internal pair 
structure between electron and hole doping is shown in Fig.~\ref{fig11}.
Here, we show the probability $C_{e}(m)$ [$C_{h}(m)$]
of finding the second doped electron
(hole) at a Cu site (in a $\text{O}_4$ orbital) when the first 
doped electron (hole) is located
on the Cu site (in the $\text{O}_4$ orbital) marked by an open circle.
The electron pair appears to be denser than the hole pair
which is qualitatively consistent with the larger binding 
energy~(\ref{bindingenergy})
calculated for the electron-doped case.

As we use open boundary conditions, the correlation functions
$C^n_{e,h}(m)$ depend on the position (i.e., the index $n$)
of the first doped carrier (electron or hole).
We have checked the pair structure for different ladder sizes
and at different positions on the ladder and have found 
significant variations close to the ladder ends. 
In Fig.~\ref{fig11} we show the pair structure in the
middle of a ladder with $32 \times 2$ Cu sites as finite-size and 
boundary effects are minimal there. 
We believe this structure to be representative of the pair structure
in an infinite ladder.

\begin{figure}
\includegraphics[width=6cm]{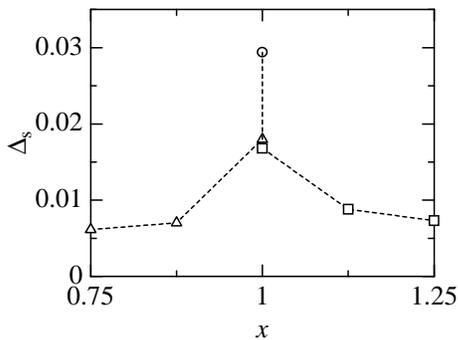}
\caption{\label{fig12}
The $L\to\infty$ extrapolated spin gap $\Delta_s$ versus hole 
concentration $x$.
The open squares denote hole doping while the triangles denote electron
doping. The undoped $x=1$ case is shown as the open circle.}
\end{figure}

As in the undoped ladder, the lowest triplet excitation in the doped
ladder is gapped and involves only holes on the Cu sites.
The spin gap~(\ref{spingap}) for the doped system
versus $x$ is shown in Fig.~\ref{fig12}. Here, we have extrapolated
the spin gap $\Delta_s(L)$ to $L\to\infty$ for fixed
hole concentrations $x = 0, \pm1/8$, and $\pm 1/4$
using numerical results for up to $L=32$. 
For the limit $x \rightarrow 1$ we have extrapolated  
the spin gap $\Delta_s(L)$ to $L\to\infty$ for the 2-hole 
and the 2-electron 
doped ladders with up to $L=32$. One sees that the value of the 
spin gap in the $x\rightarrow 1$ limit differs from the $x=1$
undoped value. 
That is, the $q=(\pi, \pi)$
magnon in the undoped system $(x=1)$
has a gap of order 0.03, while in the doped 
system the spin gap as $x \rightarrow 1$ is
of order 0.018.  This behavior is
similar to what has been found previously in
studies of $t$-$t^\prime$-$J$ ladders~\cite{Poi00}.

\begin{figure}[b]
\includegraphics[width=7.2cm]{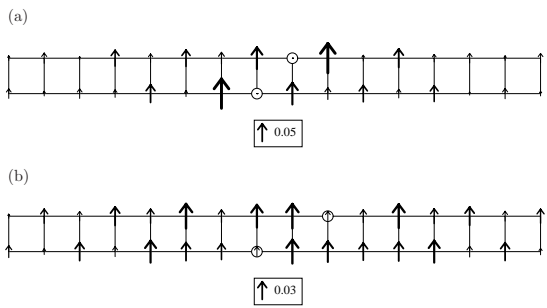}
\caption{\label{fig13}
A schematic view of a magnon in (a) an electron-doped ladder
and (b) a hole-doped ladder. 
Here, both doped carriers (electrons or holes) are located at the
positions (Cu sites or $\text{O}_4$ orbitals)
marked by open circles around the middle of a
ladder with $16 \times 2$ and $25 \times 2$ Cu sites, respectively.
The length of the arrows is proportional to the spin density 
on the corresponding Cu site.
}
\end{figure}

An examination of the spin structure of the $S_z=1$ 
state containing two doped electrons or holes shows that the spin and 
added
carriers are spatially correlated. Thus, the spin gap is set by the
energy of a bound magnon-carrier pair.
For instance, Figure~\ref{fig13} shows the spin density (on the Cu 
sites) of the $S_z=1$ state with two doped carriers (electrons or 
holes), when these doped carriers are in the configuration with the 
highest probability $C_{e,h}(m)$ (the pair structure of the triplet 
states is similar to that of the singlet states shown in 
Fig.~\ref{fig11}). 
For electron doping, the spin structure shown in Fig.~\ref{fig13}
is very similar to that found in $t$-$t^\prime$-$J$ two-leg 
ladders.~\cite{Poi00}  
For hole doping, however, the spin structure is quite different,
which is only in part due to the difference in the carrier 
configuration.
Even if we use the same carrier configuration as for the 
electron-doped case (i.e., two doped carriers on nearest-neighbor 
diagonal sites), the spin structure of the hole-doped system
is different from that found in the electron-doped system 
or $t$-$J$ model.
In particular, we note that there is a small spin density on the
Cu sites in the center of the $\text{O}_4$ orbitals where the
holes are located, which shows that in the magnon-carrier pair, 
holes on Cu and O sites are no 
longer completely locked in a Zhang-Rice singlet. 

\section{Conclusion \label{sec:conclusion}}

Using the density matrix renormalization group, we have studied the
structure of the charge, spin, and pairing correlations for a Hubbard-like
model of a two-leg CuO ladder.  This model allowed us to examine the
differences between electron and hole doping that occur in a charge
transfer insulator.  For the undoped ladder, where there is one hole per
Cu, we found that parameters obtained from LDA downfolding 
calculations~\cite{ALJP95,Mul98}
lead to reasonable charge and spin gaps.  A typical parameter set in units
of the Cu-O hopping $t_{pd}$ were a near-neighbor oxygen-oxygen hopping
$t_{pp}=0.5$, an oxygen-copper site energy difference $\Delta_{pd}=3$, and
on-site Cu and O Coulomb repulsions $U_d=8$ and $U_p=3$ respectively.
With these parameters, the charge gap is determined primarily by
$\Delta_{pd}$ rather than $U_d$ and the spin gap is set by the effective
Cu-Cu exchange.
We have found that the O-O hopping $t_{pp}$ plays a significant
role in giving a large effective Cu-Cu exchange interaction. 

For the undoped ladder, we found that the holes were distributed
approximately 70\% on the Cu sites and 30\% on the O sites. The large
on-site Coulomb interaction on the Cu site leads to the spin moment being
dominantly on the Cu site.  Basically, if the hole is on a Cu site, the
square of the  
spin moment (0.7) has nearly its full value of 3/4. When
electrons are added, they go primarily onto the Cu sites (of order 80\% of
the added charge goes onto the Cu sites). This is seen in both the decrease
in
the average Cu site hole occupation and the decrease in the average spin
moment.  Note that the decrease in the spin moment simply reflects the fact
that there are fewer Cu$^{++}$ sites in the electron-doped system.  
Alternatively, when
holes are added (electrons are removed), they go primarily onto the O sites
(of order 20\% of the holes go 
onto the Cu with 80\% going onto the
surrounding O sites). This leaves the local spin moments on the Cu sites.

Thus, the undoped CuO ladder is a spin-gapped  insulator with a charge gap
set by the oxygen-copper site energy difference. Magnetically the undoped
ladder is essentially a Heisenberg ladder made up of hole spins localized
on the Cu sites.  When the ladder is electron doped, some of the holes (and
spins) on the Cu sites are removed and pairing correlations develop.
Because the doped electrons go primarily onto the Cu sites, the
electron-doped system is closer to the one-band Hubbard or t-J models. 
When the ladder is hole doped, the holes go dominantly onto the O sites
leading to a low-density gas of fermions delocalized over the O-site
lattice and interacting by spin-exchange with the localized spins on the Cu
sites.  The local magnetic moments remain essentially unchanged on the Cu
sites but the O holes frustrate the exchange coupling between the spins on 
the Cu sites.

This local, strong coupling picture focuses on the differences between the
electron and hole-doped ladders. Nevertheless, on low energy scales, the
spin gap and 
pairing correlations of the electron- and hole-doped systems are quite
similar.  Both exhibit $d$-wave like power law pairing correlations in
which the pair field Cu-Cu rung-leg correlations are negative.  The spin
gap for both dopings is associated with a bound magnon-pair. However, the
internal structure of the pairs differ for electron and hole doping.

\acknowledgments

We would like to acknowledge useful discussions with O.K.~Anderson,
R.~Martin, and G.~Sawasky. DJS would like to acknowledge support under 
a Department of Energy Grant No.~DE-FG03-85ER-45197.

\end{document}